\documentclass[aps,prb,twocolumn,superscriptaddress]{revtex4}
\usepackage{ae}
\usepackage[T1]{fontenc}
\usepackage[ansinew]{inputenc}
\usepackage{amsmath}
\usepackage{amssymb}
\usepackage{graphicx}
\usepackage{color}

\usepackage[colorlinks]{hyperref}
\usepackage{epstopdf}

\usepackage{soul,xcolor}
\setstcolor{blue}                   % Change color of strike-through line created by "\st" here.

\begin{document}

\date{\today}

\title{Comment on: ``Spin-orbit interaction and spin selectivity for tunneling electron transfer in DNA"
}

\author{Ora Entin-Wohlman}
\email{orawohlman@gmail.com}
\affiliation{School of Physics and Astronomy, Tel Aviv University, Tel Aviv 69978, Israel}

\author{Amnon Aharony}
\email{aaharonyaa@gmail.com}
\affiliation{School of Physics and Astronomy, Tel Aviv University, Tel Aviv 69978, Israel}

\author{Yasuhiro Utsumi}
\address{Department of Physics Engineering, Faculty of Engineering, Mie University, Tsu, Mie, 514-8507, Japan}

%%%%%%%%%%%%%%%%%%%%%%%%%%%%%%%%%%%%%%%%%%%%%%%%%%%%%%%%%%%%
%%%%%%%%%%%%%%%%%%%%%%%%%%%%%%%%%%%%%%%%%%%%%%%%%%%%%%%%%%%%

\begin{abstract}
The observation of chiral-induced spin selectivity (CISS) in biological molecules still awaits a full theoretical explanation. In a recent Rapid Communication, Varela {\it et al.} [Phys. Rev. B {\bf 101}, 241410(R) (2020)]  presented a model for electron transport in biological molecules by tunneling in the presence of spin-orbit interactions. They then claimed that their model  produces a strong spin asymmetry due to the intrinsic atomic spin-orbit strength. As their Hamiltonian is time-reversal symmetric, this result contradicts a theorem by Bardarson [J. Phys. A: Math. Theor. {\bf 41}, 405203 (2008)], which states that such a Hamiltonian cannot generate a spin asymmetry for tunneling between two terminals (in which there are only a spin-up and a spin-down channels). Here we solve the model proposed by Varela {\it et al.} and show that  it does not yield any spin asymmetry, and therefore cannot explain the observed  CISS effect.
\end{abstract}

\maketitle

In spite of many theoretical papers, the observation of a large spin filtering in chiral molecules~\cite{naaman}, 
termed ``chiral induced spin selectivity (CISS)", still awaits a full explanation, which is accepted by everyone.
In a recent Rapid Communication, Varela {\it et al.}~\cite{Varela} followed a series of their earlier papers, and mapped the detailed tunneling electron transfer through the molecule onto an effective one-dimensional continuum model, which mimics the molecule by a region with a barrier potential and a Rashba spin-orbit interaction (SOI). Using a scattering solution of this model, they concluded that the molecule causes spin-splitting of the scattered electrons, thus explaining the CISS experiments.

Since the Rashba SOI obeys time-reversal symmetry, the above result contradicts a general theorem by Bardarson~\cite{Bardarson}, which states that  a time-reversal symmetric Hamiltonian cannot generate a spin asymmetry for tunneling between two terminals (in which there are only a spin-up and a spin-down channels)~\cite{r1}. Indeed, this led several groups to propose models which effectively break time-reversal symmetry without a magnetic field for two-terminal systems~\cite{us1}, or to increase the number of channels~\cite{us2}. Below we solve the model of Ref. \onlinecite{Varela} explicitly, and show that indeed  it does not generate any spin splitting, thus obeying the Bardarson theorem.

After several mappings, Ref.~\onlinecite{Varela} ends up with a one-dimensional Hamiltonian for the electronic spinors on the molecule, Eq. (5) in 
that  paper,
\begin{align}
{\cal H}=\Big[\frac{p_x^2}{2m}+V^{}_0\Big]{\bf 1}+\alpha \sigma^{}_y p^{}_x\ \ \ {\rm for}\ \ 0<x<a\ ,
\label{HH}
\end{align}
where $a$ is the molecule's length, $\sigma^{}_y$ is the Pauli spin matrix, ${\bf 1}$ is the $2\times 2$ unit matrix, $\alpha$ represents the strength of the spin-orbit interaction, and $V^{}_0$ represents an energy barrier on the molecule. For $x<0$  and $a<x$ Ref. \onlinecite{Varela} has $V^{}_0=0$ and $\alpha=0$, and therefore the Hamiltonian in those regions is that of free electrons, $p_x^2/(2m)$, with arbitrary spinors, with a spatial wave function $e^{\pm i k x}$, and  energy $E=\hbar^2k^2/(2m)$.

 It is convenient to choose as a basis of the spin Hilbert space the eigenspinors of $\sigma^{}_y$, $\sigma^{}_y|\mu\rangle=\mu|\mu\rangle$, with $\mu=\pm 1$, and write the solutions as $|\Psi^{}_\mu(x)\rangle=\psi^{}_\mu(x)|\mu\rangle$.
Applying ${\cal H}$ to each of these states yields
\begin{align}
{\cal H}|\Psi^{}_\mu(x)\rangle=\Big[\frac{p_x^2}{2m}+V^{}_0+\alpha\mu p^{}_x\Big]|\Psi^{}_\mu(x)\rangle\ .
\end{align}
In the chosen basis, the Hamiltonian is diagonal, and this equation separates into two scalar equations. In the range $0<x<a$ these are
\begin{align}
\Big[\frac{p_x^2}{2m}+V^{}_0+\alpha\mu p^{}_x\Big]\psi^{}_\mu(x)=E\psi^{}_\mu(x)\ .
\label{psii}
\end{align}
Assuming a solution of the form $\psi^{}_\mu(x)\propto e^{iQ^{}_\mu x}$, we find that $Q^{}_\mu$ must
 obey the quadratic equation
\begin{align}
E=\frac{\hbar^2[( Q^{}_\mu+k^{}_{\rm so} \mu)^2-k_{\rm so}^2]}{2m}+V^{}_0\ ,
\label{EE}
\end{align}
where $m\alpha/\hbar=k^{}_{\rm so}$ is the strength of the SOI in units of inverse length. This equation has two solutions,
\begin{align}
Q^{\pm}_\mu=-k^{}_{\rm so}\mu\pm q\ ,\ \    {\rm with}\ \ q=\sqrt{k^2+k_{\rm so}^2-q_0^2}\ ,
\label{our}
\end{align}
where $q_0^2=2mV^{}_0/\hbar^2$.

Our Eq. (\ref{our}) differs from Eq. (7) of Ref. \onlinecite{Varela}, which in our notation would be:
\begin{align}
Q^{\pm}_\mu({\rm Varela})=\pm(k^{}_{\rm so}\mu+q)\ .
\label{VAR}
\end{align}
Clearly these values do not obey  Eq. (5) of Ref. \onlinecite{Varela} [and  our Eq. (\ref{EE})]. We suspect that this discrepancy led to the spin splitting found there.

Explicitly, one faces  a simple scattering problem,~\cite{com}
\begin{align}
\psi^{}_\mu&=\big[e^{i k x}+r^{}_\mu e^{-i k x}\big]\ ,\ \ \ x<0\ ,\nonumber\\
\psi^{}_\mu&=e^{-ik^{}_{\rm so}\mu x}\big[C^{}_\mu e^{i q x}+D^{}_\mu e^{-iq x}\big]\ ,\ \ \ 0<x<a\,\nonumber\\
\psi^{}_\mu&=t^{}_\mu e^{i k x}\ ,\ \ \ a<x\ .
\label{psii}
\end{align}
The prefactor in the middle region is nothing but the  Aharonov-Casher phase factor~\cite{AC} due to the spin-orbit interaction. The SOI adds opposite phases to the two spin states.

Generally, the conjugate velocity is given by $v=\partial{\cal H}/(\partial p^{}_x)$. For each of the four solutions in Eq. (\ref{our}),  the corresponding gauge covariant velocities inside the molecule are $v^{\pm}_\mu=\hbar(Q^{\pm}_\mu+k^{}_{\rm so}\mu)/m=\pm \hbar q/m$. For $E>V_0-(\hbar k^{}_{\rm so})^2/(2m)$, $q$ is real, and the solution on the molecule has waves propagating to the right and to the left. For $E<V_0-(\hbar k^{}_{\rm so})^2/(2m)$, $q$ is imaginary, and the waves become evanescent. The continuity conditions at $x=0$ and $x=a$ yield  four equations for the four unknowns $C^{}_\mu,~D^{}_\mu,~r^{}_\mu$ and $t^{}_\mu$:
\begin{align}
1+r^{}_\mu&=C^{}_\mu+D^{}_\mu\ ,\nonumber\\
k(1-r^{}_\mu)&=q\big[C^{}_\mu-D^{}_\mu\big]\ ,\nonumber\\
t^{}_\mu e^{i k a}&=e^{-ik^{}_{\rm so}\mu a}\big[C^{}_\mu e^{i q a}+D^{}_\mu e^{-iq a}]\ ,\nonumber\\
k t^{}_\mu e^{i k a}&=q e^{-ik^{}_{\rm so}\mu a}\big[C^{}_\mu e^{i q a}-D^{}_\mu e^{-iq a}\big]\ .
\label{5}
\end{align}
Replacing $t^{}_\mu$ by $\tilde{t}^{}_\mu=t^{}_\mu e^{ik^{}_{\rm so}\mu a}$ yields equations which are independent of $\mu$, and therefore the solutions for $r^{}_\mu$ and $\tilde{t}^{}_\mu$ are independent of $\mu$. Since the transmission and reflection probabilities are $T^{}_\mu=|t^{}_\mu|^2=|\tilde{t}^{}_\mu|^2$ and $R=|r^{}_\mu|^2$, it is clear that the reflection and transmission matrices $R$ and  $T$ are proportional to the $2\times 2$ unit matrix, and therefore there is no spin selection, in accordance with the Bardarson theorem~\cite{Bardarson}. The model of Ref. \onlinecite{Varela} does not generate any asymmetry in the outgoing spin currents.

Specifically, the solutions are
\begin{align}
r^{}_\mu&=\frac{k^2-q^2}{q^2+ k^2 +
 2 i k q \cot(qa)}\ ,\nonumber\\
 t^{}_\mu&=\frac{2 e^{-i a (k^{}_{\rm so}\mu  + k)} k q}{
2 k q \cos(qa) -
 i (k^2+q^2) \sin(qa)}\ .
 \label{rt}
 \end{align}
 and thus
 \begin{align}
T^{}_\mu=|t^{}_\mu|^2=\frac{4k^2q^2}{4k^2q^2+\big(k^2-q^2)^2\sin^2(qa)}\ ,
\end{align}
 {\bf independent of $\mu$!} It is also straightforward to check unitarity, $R^{}_\mu+T^{}_\mu=1$.  This result also holds when $q$ is purely imaginary.
Solving the same equations with the $Q$'s used in Ref. \onlinecite{Varela}, Eq. (\ref{VAR}), indeed yields different velocities for the two spins, ending up with spin-dependent reflection and transmission.

%Another subtle problem with the Hamiltonian (\ref{HH}) is that it is not hermitian: ${\cal H}\ne{\cal H}^\dagger$, due to the abrupt changes of %the  SOI coefficient $\alpha$ at the interfaces ~\cite{morpurgo}. The `standard' way to overcome such difficulties, replacing ${\cal H}$ by %$({\cal H}+{\cal H}^\dagger)/2$, does not work, since the latter Hamiltonian breaks time-reversal symmetry.

An alternative way to derive the scattering amplitude is to first apply a gauge transformation (related to the Aharonov-Casher phase factor~\cite{AC}),
\begin{align}
|\Psi(x)\rangle=U(x)|\widetilde{\Psi}(x)\rangle\ ,\ \ \ U(x)=e^{-i k^{}_{\rm so}x\sigma^{}_y}\ ,
\label{UU}
\end{align}
so that
\begin{align}
\widetilde{\cal H}=U(x)^\dagger{\cal H}U(x)=\frac{p_x^2-(\hbar k_{\rm so})^2}{2m}+V^{}_0\ .
\end{align}
This is a spin-independent hermitian Hamiltonian, whose eigenstate in the `molecule' region has the form
\begin{align}
\widetilde{\psi}(x)=\widetilde{C}e^{iqx}|+\rangle+\widetilde{D}e^{-i q x}|-\rangle\ ,
\label{tilpsi}
\end{align}
with the same $q=\sqrt{k^2+k_{\rm so}^2-q_0^2}$ given in Eq. (\ref{our}).
The boundary conditions for $\widetilde{\psi}$ are the same as for spinless particles, hence the transmission amplitude is
\begin{align}
 \widetilde{t}=\frac{2 e^{-i a k} k q}{2 k q \cos(qa) - i (k^2+q^2) \sin(qa)}\ .
 \end{align}

From Eq. (\ref{UU}), $|\Psi(a)\rangle=U^\dagger(a)|\widetilde{\Psi}(a)\rangle$.
Noting that $U(x)|\pm\rangle=e^{\mp i k^{}_{\rm so}x}|\pm\rangle$, it follows that $t^{}_\mu=e^{-i a k^{}_{\rm so}\mu}\widetilde{t}^{}_\mu$, reproducing Eq. (\ref{rt}) and the spin-independence of the transmission probability.
In fact, the gauge transformation simply shifts  the covariant momentum $\tilde{p}^{}_x=p^{}_x+\hbar k^{}_{\rm so}\mu$ onto the momentum $p^{}_x$, which is also seen directly from Eq. (\ref{EE}). This results in a simple Aharonov-Casher phase shift in the transmission amplitude, and does not affect the transmission probability. The reflection and transmission probabilities are invariant under the gauge transformation, and therefore remain spin-independent.

In conclusion, one cannot generate spin splitting  with only spin-orbit interactions, as done in Eq. (5) of Ref. \onlinecite{Varela}, and the chiral induced spin selectivity effect still awaits a full theoretical explanation.

%%%%%%%%%%%%%%%%%%%%%%%%%%%%%%%%%%%%%%%%%%%%%%%%%%%%%%%%%%%%%%%

\begin{acknowledgements}
We thank an anonymous referee for drawing our attention to Ref. \onlinecite{Varela}. We acknowledge support by JSPS KAKENHI Grants 17K05575, 18KK0385, and 20H01827, and  by the Israel Science Foundation (ISF), by the infrastructure program of Israel Ministry of Science and Technology under contract 3-11173, and  by the Pazy Foundation.
\end{acknowledgements}
%%%%%%%%%%%%%%%%%%%%%%%%%%%%%%%%%%%%%%%%%%%%%%%%%%%%%%%%%%%%%%

%%%%%%%%%%%%%%%%%%%%%%%%%%%%%%%%%%%%%%%%%%%%%%%%%%%%%%

%%%%%%%%%%%%%%%%%%%%%%%%%%%%%%%%%%%%%%%%%%%%%%%%%%%%%%


\begin{thebibliography}{99}
\bibitem{r1} Time-reversal symmetry was similarly shown earlier~\cite{r2} to cause the invariance of the linear conductance through an interface between a ferromagnet and a Rashba-active two-dimensional semiconductor under the inversion of the magnetic moment in the ferromagnet. These papers emphasize the importance of the correct treatment of the velocity operator [as discussed before Eq. (\ref{5})].
\end{thebibliography}
\end{document}